\documentclass[a4paper,UKenglish,cleveref, autoref, thm-restate]{lipics-v2021}
\usepackage{tabularx}
\usepackage{booktabs}

\usepackage{amsmath}
\DeclareMathOperator*{\argmax}{arg\,max}

\title{A Reinforcement Learning based Reset Policy for CDCL SAT Solvers} %TODO Please add

\author{Chunxiao Li}{University of Waterloo, Canada}{chunxiao.li@uwaterloo.ca}{https://orcid.org/0000-0001-7336-1614}{}
\author{Charlie Liu}{University of Waterloo, Canada}{charlie.liu@uwaterloo.ca}{}{}
\author{Jonathan Chung}{University of Waterloo, Canada}{chunxiao.li@uwaterloo.ca}{}{}
\author{Zhengyang (John) Lu}{University of Waterloo, Canada}{john.lu2@uwaterloo.ca}{https://orcid.org/0009-0005-9046-497X}{}
\author{Piyush Jha}{University of Waterloo, Canada}{piyush.jha@uwaterloo.ca}{}{}
\author{Vijay Ganesh}{University of Waterloo, Canada}{vijay.ganesh@uwaterloo.ca}{https://orcid.org/0000-0002-6029-2047}{}

\authorrunning{J. Open Access and J.\,R. Public} %TODO mandatory. First: Use abbreviated first/middle names. Second (only in severe cases): Use first author plus 'et al.'

\Copyright{Jane Open Access and Joan R. Public} %TODO mandatory, please use full first names. LIPIcs license is "CC-BY";  http://creativecommons.org/licenses/by/3.0/


 %TODO mandatory; please add comma-separated list of keywords

\category{} %optional, e.g. invited paper

\relatedversion{} %optional, e.g. full version hosted on arXiv, HAL, or other respository/website
\nolinenumbers %uncomment to disable line numbering

\EventEditors{John Q. Open and Joan R. Access}
\EventNoEds{2}
\EventLongTitle{42nd Conference on Very Important Topics (CVIT 2016)}
\EventShortTitle{CVIT 2016}
\EventAcronym{CVIT}
\EventYear{2016}
\EventDate{December 24--27, 2016}
\EventLocation{Little Whinging, United Kingdom}
\EventLogo{}
\SeriesVolume{42}
\ArticleNo{23}
\newcommand{\rwglr}{rw\_glr}

\begin{document}

\maketitle

%TODO mandatory: add short abstract of the document
\begin{abstract}
Restart policies are an important and widely studied class of techniques used in state-of-the-art Conflict-Driven Clause Learning (CDCL) Boolean SAT solvers, wherein some parts of the state of solvers is erased at certain intervals during the run of the solver. In most modern solvers, variable activities are preserved across restart boundaries. An implication of this is that solvers continue to search parts of the assignment tree that are not far from the one immediately prior to a restart. To enable the solver to search possibly ``distant'' parts of the assignment tree, we study the effect of resets, a variant of restarts which not only erases the assignment trail, but also randomizes the activity scores of the variables of the input formula after reset, thus potentially enabling a better {\it global exploration} of the search space. Just like in the case of restarts, the crucial design question here is when and how frequently to invoke reset.

In this paper, we model the problem of whether or not to trigger reset as a multi-armed bandit (MAB) problem, and propose two reinforcement learning (RL) based adaptive reset policies using the Upper Confidence Bound (UCB) method and Thompson sampling. The UCB method and Thompson sampling algorithms are designed to balance the exploration-exploitation tradeoff by probabilistically and adaptively choosing arms (reset vs. no reset) based on their estimated rewards during the solver's run. We implement our reset policies in four different state-of-the-art CDCL solvers, namely, CaDiCaL, SBVA\_Cadical, Kissat and MapleSAT, and compare the baselines against the reset versions on a set of 500 Satcoin benchmarks and 800 Main Track instances from the SAT competition 2022 and 2023. Our results show that RL-based reset versions of these solvers perform better than the corresponding baseline solvers on Satcoin and the SAT competition instances, suggesting that the adaptive aspect of our RL policy helps the solver to dynamically and profitably adapt the reset frequency for any given input instance. We also introduce the concept of a partial reset, where at least a constant number of variable activities are retained across reset boundaries. Building on previous results, we show that there is an exponential separation between $O(1)$ vs. $\Omega(n)$-length partial resets.
\end{abstract}

\section{Introduction}
Over the last few decades, Conflict-Driven Clause Learning (CDCL) solvers have had a dramatic impact on many areas of software engineering~\cite{cadar2008exe}, security~\cite{dolby2007security,xie2005security}, and AI~\cite{satplan1,satplan2}. This is in part due to heuristics for branching (e.g., VSIDS~\cite{moskewicz2001chaff} and LRB~\cite{liang2016learning}), deletion~\cite{goldberg2007berkmin}, and restart~\cite{gomes1998boosting}, to name just a few. A fundamental problem that solver designers often face is that a set of techniques that work well for a class of instances can fail miserably for another. 

Restart policies are an excellent example of the above described phenomena. While these policies may differ in terms of frequency of restart, almost all of them have one feature in common, namely, that activities of variables are not changed across restart boundaries. This works well for verification instances found in the SAT Competition Main Track benchmark suites. However, this is not true for many other classes such as cryptographic instances (we discuss this observation in the following sections). In fact, in a recently published paper on restart policies~\cite{vinyals20hard} it was shown that maintaining activities of variables across restart boundaries is the key reason as to why CDCL solvers with VSIDS-style branching heuristics and traditional restarts have exponential lower bound on a class of crafted instances called pitfall formulas. By contrast, in a follow-up paper~\cite{li2020complexitytheoretic}, it was shown that CDCL solvers can polynomially solve pitfall formulas if variables activities are assigned 0 after restart boundaries and ties are broken randomly during branching (or equivalently, variable activities are randomized across restart boundaries) while keeping the rest of the CDCL configuration the same. We refer to this variant of restart as reset~\footnote{We do not claim to be the first to come up with the idea of reset. While we know that practitioners have implemented variants of resets, we couldn't find any papers on the topic. Further, some theory papers also talk about variants of resets, including Li et al.~\cite{li2020complexitytheoretic}. What is new in our paper is the idea of using RL to determine adaptively whether or not to invoke reset.}.

We also performed some empirical observations about the performance some of the best recent SAT solvers, namely Kissat 3.0.0~\cite{fleury2020cadical},  and the latest version of MapleSAT~\cite{liang2016learning}, on cryptographic instances such as Satcoin~\cite{manthey2018satcoin}~\footnote{These instances are obtained from a bitcoin mining application that uses the SHA256 Hash function~\cite{manthey2018satcoin}.}. Specifically, we observed that both these solvers performed poorly on such instances. Upon experimenting with a variety of restart policies, we finally observed that when the activities were periodically reset to random values, the solvers' performance improved dramatically. Further, as we increased the frequency of reset, we observed that these reset-based solvers performed better on Satcoin benchmarks. By contrast, these reset variants behaved poorly on SAT competition Main Track instances, even though their non-reset cousins performed well. Finally, as we experimented with greater frequency of reset, the performance of these reset variants got progressively worse.

This observation leads us to ask ``can we get the best of both worlds, i.e., develop an adaptive technique that dynamically and optimally switches between restarts and resets during the run of a solver for any given instance such that it outperforms, in terms of solving time, any other possible ordering of restart and reset calls?''

Based on the above mentioned observation, we propose two reinforcement learning (RL) based policies that enables the solver to switch between restart and reset policies adaptively (and hopefully optimally) during the solver's run, using the Thompson sampling and UCB algorithms. The intuition is based on the idea that RL-based policies can enable solvers to adapt to any given instance during a solver's run time, by dynamically selecting variables (a la the LRB branching heuristic~\cite{liang2016learning}) or switching between heuristics~\cite{lagoudakis2001learning, cherif2021combining}, and thus be more efficient than if it was designed to use non-adaptive policies.

More precisely, we model the problem of whether a solver should perform a traditional restart or a reset as an MAB problem. There are two arms in our model: pulling the first arm corresponds to performing a traditional restart, while pulling the second arm corresponds to performing a reset. Each time a restart is supposed to happen in a CDCL solver, we call our RL model and ask it whether we should pull arm one or arm two. To identify and reward the action of pulling each arm, we came up with a reward mechanism based on the notion of learning rate (LR), which is a well studied empirical measure of efficiency used in SAT solving. We showcase the efficacy of our techniques via extensive experiments over multiple state-of-the-art solvers. We also contrast our work against previous theoretical and empirical work on reset policies. 

\vspace{0.2cm}
\noindent{\bf Contributions.}

\begin{enumerate}
    \item {\bf RL-based adaptive reset-invocation techniques:} First, we model the problem of deciding whether to reset as a multi-armed bandit (MAB) problem, and we solve the problem using the Upper Confidence Bound (UCB) method and Thompson sampling, two well-known RL methods. This gives the solver a chance to adaptively decide when to choose an arm, by using the observed successes and failures from previous pulls of the arms during the run of the solver on any given instance. In order to adapt the Thompson Sampling algorithm to the context of CDCL solvers, where the underlying search space changes may get restricted over time due to clause learning, and then expand due to clause deletion, we came up with a novel technique which decays the shape parameters used in the Thompson Sampling algorithm.
    % John: I will add UCB variant for non-stationary here; also may rewrite the decay justification for Thompson

    \item{{\bf Partial resets:} Second, we propose a technique that is a variant of resets, which we call partial resets. As opposed to randomly shuffling the activity scores for all variables, a partial reset preserves the order of top activity variables prior to reset boundaries. That is, the top $k$ after a reset is exactly the set of variables with top $k$ activities in the exact same order. This technique aims at preserving some ``locality'' information of the branching heuristic while re-initializing the search after a reset. We also observe that from a theoretical point of view, partial reset lies in between restart and full reset. That is, we show that there is an exponential separation between $O(1)$ vs. $\Omega(n)$-length partial resets.}
    
    \item {\bf Extensive empirical evaluation and ablation studies:} Lastly, we present empirical evaluations of our RL reset policy on three sets of benchmarks (SAT Competition Main Track 2022/2023 and Satcoin) over the following state-of-the-art baseline solvers~\footnote{https://github.com/ChunxiaoIanLi/RL\_reset}, namely, Kissat~\cite{fleury2020cadical}, MapleSAT~\cite{liang2016learning}, CaDiCaL~\cite{BiereFazekasFleuryHeisinger-SAT-Competition-2020-solvers}, SBVA\_Cadical~\cite{haberlandt2023effective}. We show that the modified versions of these solvers with our RL based reset policy performs better or as well than the corresponding baselines, while retaining their competitiveness against the baselines on SAT Competition 2022 and 2023 Main Track instances~\cite{SATcomp}.
\end{enumerate}
\section{Preliminaries} 

\subsection{Restart Policies}
A restart policy is a method that erases part of the state\footnote{Roughly speaking, the state of a CDCL SAT solver can be defined to include the assignment trail, variable activities and phases, as well as the clause databases.} of the solver at certain intervals during the run of a solver. In most modern CDCL solvers, the restart policy erases the assignment trail upon invocation, but does not erase the learnt clause database and variable activities. We refer the reader to a paper by Liang et al.~\cite{liang2018machine} for a detailed discussion on most widely used restart policies, and a paper by Li et al.~\cite{li2020complexitytheoretic} for theoretical discussions on restart policies in modern CDCL solvers.

To overcome the limitation of biased ``local'' search introduced by restarts, the concept of resetting has been proposed, where the activity scores of the variables are randomized after the assignment trail is erased. Traditionally, a reset refers to zeroing out or randomizing activity scores of all variables of an input formula. Below we provide formal definitions for reset strategies, categorized as full and partial reset.

\begin{definition}[Full Reset]
    Upon invocation, a full reset policy randomizes the activity scores of all variables (and thus randomizes the variable order for branching), in addition to deleting the contents of the assignment trail as in restart policies.
\end{definition}

\begin{definition}[($k$-)Partial Reset]
    Upon invocation, a partial reset policy retains the top $k$ variables in the same exact order as before the reset, and the order of all the remaining variables is randomized. The contents of the assignment trail are deleted.
\end{definition}

%\subsection{Reinforcement Learning}
%Reinforcement Learning (RL) is an area of machine learning that focuses on training agents to take actions (make decisions) in an environment in order to maximize some notion of reward. The agent learns through a process of interacting with the environment and receiving feedback in the form of rewards or penalties.
% RL frameworks are typically based on the idea of a Markov Decision Process (MDP), which is a mathematical model for decision making in which the outcomes of actions are probabilistic and depend on the current state of the environment. The agent's objective is to learn a policy, which is a mapping from environment states to actions, that maximizes the expected cumulative reward over time. 

\subsection{RL and the Multi-Armed Bandit (MAB)}
For a detailed discussion on RL and the Multi-Armed Bandit (MAB) problem, we refer the reader to the Sutton book~\cite{sutton2018reinforcement}. Here we provide a brief overview of the MAB problem for completeness. 

The MAB problem is one of the simplest RL problems and is particularly useful when we want to model an agent that has to make frequent actions in an uncertain environment. In MAB problems, there is a given set of slot machines with unknown payout distributions, and a solution to the problem corresponds to a strategy for playing the slot machines which minimizes regret. To put it in RL terms, the agent, i.e., the player (gambler), frequently takes actions of selecting an arm to pull and receives feedback from the environment, i.e., the slot machine. Going through this feedback loop repeatedly, the agent learns which arm is more likely to provide the highest reward and selects it more frequently.

An effective MAB strategy requires a trade-off between exploration and exploitation. Selecting a less explored arm to play is considered exploration, while repeatedly playing a previously rewarding arm is considered exploitation. After a sufficient amount of exploration, an RL agent may have enough confidence in the expected rewards from different arms that it can start exploiting the arms with the highest expected reward. 

One important and relevant topic in the context of MABs is how to deal with non-stationary environments, i.e., the reward distribution of each arm may change over time. Most effective non-stationary MAB algorithms assign more weight to recent rewards than to long-past rewards when estimating each arm's expected reward, e.g., via the Exponential Recency Weighted Average (ERWA) method~\cite{sutton2018reinforcement}. Such considerations will be discussed when we present our reset algorithms.

\subsection{Thompson Sampling}
Thompson sampling~\cite{thompson1933likelihood}, also called Probability Matching Strategies or Bayesian Bandits is a well-known approach for solving the MAB problem. In Thompson sampling, the method initializes a model distribution for each arm, and then updates each distribution based on the cumulative history of the outcome of pulling the corresponding arm. It is designed to balance the exploration-exploitation tradeoff by probabilistically choosing arms based on their estimated rewards. The algorithm works by sampling a reward from the posterior distribution of each arm, and selecting the arm with the highest sample.

In Thompson Sampling, the algorithm assumes that the prior distribution of each arm's reward is a beta distribution. The beta distribution is a continuous probability distribution that takes values between 0 and 1. The beta distribution has two shape parameters, denoted by $\alpha$ and $\beta$, which control the shape of the distribution. The mean of the distribution is given by $\alpha/(\alpha + \beta)$, and the variance is given by $\alpha \beta/((\alpha+\beta)^2(\alpha+\beta+1))$. In the context of Thompson Sampling, the shape parameters $\alpha$ and $\beta$ are often used to represent the number of successes and failures of each arm, respectively. Typically, we would consider pulling an arm as a success if the reward from pulling the arm is ``good enough''. What is ``good enough'' depends on the context and should be customized to adapt to the underlying problem.

Each time an arm is pulled, the algorithm updates the shape parameters for the beta distribution of that arm based on the outcome of the pull. If the pull results in a success, $\alpha$ for that arm is incremented by 1, and if it results in a failure, the arm's $\beta$ is incremented by 1. These updates help to refine the algorithm's estimate of the reward distribution for each arm, which in turn informs its future decisions about which arm to pull. By using the beta distribution in this way, Thompson Sampling can balance the exploration-exploitation tradeoff by probabilistically selecting arms based on their estimated reward distributions. This approach allows the algorithm to explore arms with uncertain rewards while also exploiting arms with high estimated rewards, often leading to better performance.

\subsection{Upper Confidence Bound (UCB)}
Another common and effective method for addressing the exploration-exploitation dilemma in MAB problems is the UCB1 algorithm \cite{ucb1_2002}, which selects the action $A_t$ at each time step $t$ according to:

$$A_t = \underset{a}{\argmax}[Q_t(a)+c\sqrt{\frac{\ln t}{N_t(a)}}]$$

\noindent where $Q_t(a)$ denotes the estimated reward value of action $a$ at time step $t$, $N_t(a)$ denotes the number of times that action $a$ has been selected prior to time $t$, and the constant $c > 0$ controls the degree of exploration. The max'ed sum (UCB value) represents an estimated upper bound of each $a$'s true reward, within which the second term is a measure of uncertainty in the reward estimation. To tackle the environment non-stationariness, Garivier and Moulines \cite{swucb_2011} introduce UCB variants, i.e., discounted UCB (D-UCB) and sliding-window UCB (SW-UCB).

\begin{figure}[t]
\centering
{\includegraphics[width=1\linewidth]{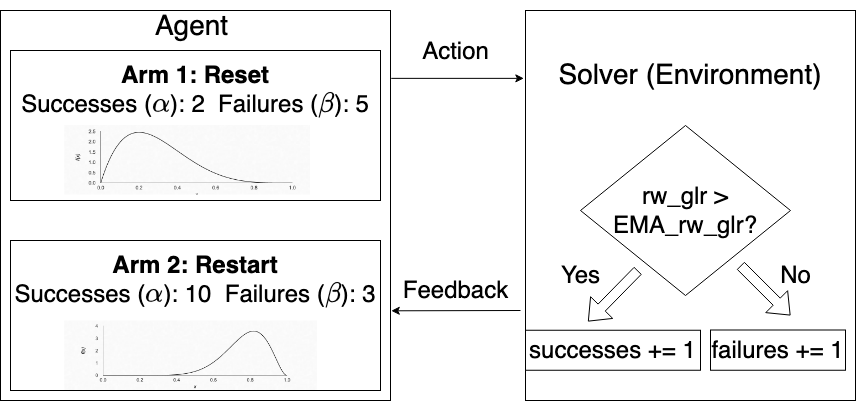}}
\caption{The figure describes our reset method. Immediately prior to calling reset (and after the assignment trail has been deleted), we sample from the beta distribution for each arm and pick the arm with the higher of the two sampled value. Further, we then compute the \rwglr{} after the action is taken, update the EMA\_\rwglr{}, and compare the \rwglr{} and EMA\_\rwglr{}. If \rwglr{} $>$ EMA\_\rwglr{}, the number of success is increased by 1 for the selected arm, and if \rwglr{} $\le$ EMA\_\rwglr{}, the number of failure is increased by 1.}
\label{fig:rl_flow} 
\end{figure}

\section{Reset Models}
\label{sec:reset_models}

The problem of whether to restart or reset in an SAT solver can be modeled as an RL problem, where a crucial aspect is the design of rewards. Here, we define a metric {\it restart window global learning rate (\rwglr{})} as below:

% by appropriately defining the states, actions, rewards, and learning objective, similar to previous work on RL for SAT heuristics such as RL-based branching~\cite{liang2016learning} as well as restarts~\cite{liang2018machine}. 

% ``Local'' here refers to the solver's performance (via the notion of learning rate) in between two consecutive restart boundaries. 
% We then use llr to evaluate whether the previous pull of arms is a success or failure. 

\begin{definition}[restart window local learning rate({\it \rwglr{}})]
\end{definition}

% John: capitalize or italicize llm

% John: probably put reference after different metrics
Traditionally, a range of metrics including solving time, number of decisions, number of conflicts, average width and many others have been used to evaluate SAT solver's performance. In this paper, we choose to use {\it \rwglr{}}, which offers a new and meaningful dimension for assessing SAT solver effectiveness. Different solvers have varied branching behaviours, learning techniques, clause deletion policies, making it challenging to compare them in an apple-to-apple fashion. The {\it \rwglr{}} metric gives a normalized measure of how effective a solver is at generating learned clauses for every decision it makes. If a solver learns more clauses with the same number of decisions, it means that the solver is more efficient in exploring its current search space. This efficiency could lead to faster solving time. 

Using the \textit{\rwglr{}} as the reward, the RL problem of reset can be modeled as the environment state corresponding to the SAT solver state (e.g., assignment trail, learnt clauses etc.) and the action corresponding to performing a restart or a reset.
However, attempts to model the online heuristic selection in SAT solvers as a full RL problem, aimed at learning a policy that maps solver states to decisions, often lead to sub-optimal performance in modern solvers \cite{GQSAT2019, mcfs_2022} due to significant computational overhead. The primary challenge lies in the necessity to employ complex neural networks for policy or value function approximation. This necessity arises from the need to navigate the solver's state space, which is both vast and intricate. These neural network models, due to their complexity, introduce significant computational overheads. This additional burden may undermine the efficiency of modern SAT solvers. 

One successful alternative is to model such problems as a non-stationary MAB, as shown by the work of Liang et al. in RL-based branching~\cite{liang2016learning} and RL-based restarts~\cite{liang2018machine}. In such modeling, the solver states are not explicitly considered while the changing environment is captured by the non-stationary assumption. This allows us get the benefit of an RL-based heuristic without paying the high computational overhead of stateful RL techniques. Our work follows this trend and proposes various models to solve the reset MAB problem.

For the rest of this section, we describe five reset models we study in this paper. 

\subsection{Full Reset with Fixed Probabilities}
In the reset policy with fixed probabilities, the reset is performed with a predetermined probability $p$, at each restart boundary. This model serves as a baseline for exploring the potential of adaptive resets. 

\subsection{Full Reset with Thompson Sampling}
For this method, we adapt the Thompson sampling technique for solving our MAB model of the reset problem. We do this by first defining what is meant by success and failure. Ideally, we would like to count as success those actions that lead to effective pruning of the search space. There have been various metrics proposed and studied to measure progress, including metrics to capture the quality of learnt clauses, decisions etc. For example, the size of clauses and literal block distance (LBD)~\cite{audemard2009predicting} are well-known metrics to measure the quality of learnt clauses. Activities and global learning rate are examples of quality metrics for branching heuristics~\cite{moskewicz2001chaff, liang2017empirical}. 

We chose the notion of global learning rate to evaluate the performance of the arms in our model because it is well known to be an effective measure of a solver's progress~\cite{liang2016learning}.
To be more specific, we use an \textit{exponential moving average (EMA)} of restart window global learning rate, which we refer to as EMA\_\rwglr{}, to keep track of each arm's historical performance, and then consider the pulling of an arm at $t_i$ as a success if EMA\_\rwglr{} $>$ \rwglr{}$(i, i+1)$. Otherwise when EMA\_\rwglr{} $\le$ \rwglr{}$(i, i+1)$, we would increase the failure count by one. Figure~\ref{fig:rl_flow} contains the overall structure of reset model using Thompson Sampling.

\subsection{Full Reset with Thompson Sampling and Decaying Shape Parameters} In the context of whether to restart or reset, the output of the Thompson sampling algorithm relies on the historical performance of resetting. However during a solver's run, the solver keeps learning conflict clauses as well as deleting useless clauses over time, which means the underlying search space gradually changes all the time. Thus a success (or a failure) for an arm at the beginning of the run may be quite useless for determining which arm to pull at a later stage of the solve run. To solve this problem, we apply a decay, $0 < d < 1$, to the shape parameters every time one of the shape parameters gets updated, to give more weights to successes and failures of the arms' recent performances. Without loss of generality, assume at the time we would like to increase $\alpha$ by one, instead of just adding one to $\alpha$, we do $\alpha_{new}=\alpha*d + 1$, and at the same time, we update $\beta$ as well by doing $\beta_{new}=\beta*d$. Through our experiments, we found that decaying shape parameters are crucial to the performance of our RL reset policy. 

\subsection{Full Reset with SW-UCB}
We use the Sliding-Window UCB (SW-UCB) algorithm for the non-stationary reset MAB problem. Unlike the traditional UCB, which considers the entire history of actions and rewards, SW-UCB focuses on a recent window of observations to adapt to changes in the reward distributions of actions. Our SW-UCB algorithm selects actions based on the UCB value within a fixed-size sliding window that moves over time. The window size $\tau$ is predetermined. 

\subsection{Partial resets} 

We also explore a variant of full reset, we refer to as partial reset. As defined earlier, in a partial reset the activities of the top-$k$ variables are retained across reset boundaries and the activities of the remaining variables are randomized.To achieve this, the ordering of top $k$ variables are recorded before randomizing activities, and after randomizing activities for all variables, the top $k$ variables receive a constant bump, just enough so that they still have top activities, and in the same order as before the reset.  The reason for exploring partial resets is that full reset may not be the best strategy for industrial instances that have a lot of ``locality''.
\section{Theoretical Observations}

Below we make a few theoretical observations about our reset policies. 

\subsection{Separation Results for Full and Partial Reset}

Vinyals~\cite{vinyals20hard} demonstrate the existence of a class of instances, called pitfall formulas, featuring polynomial-size resolution proofs and exponential lower bounds for CDCL solvers employing Variable State Independent Decaying Sum (VSIDS)-like branching heuristics alongside restarts.

\begin{theorem} \label{thm:vinyals20hard} {\bf Vinyals' Theorem}~\cite{vinyals20hard}:
There is a family of instances, called pitfall formulas, that have polynomial resolution proofs, but CDCL with VSIDS and restarts requires exponential time to solve them, except with exponentially small probability.
\end{theorem}

In a separate paper, Li, Vinyals, and collaborators~\cite{li2020complexitytheoretic} present polynomial size proofs of these pitfall formulas using CDCL solvers with VSIDS-like branching and a reset policy. Notably, the solver examined by Li et al. deviates in two significant aspects from those analyzed in Vinyals' theorem. Firstly, Li et al.'s solver resets the VSIDS activity scores for all variables to zero post-restart. Secondly, their VSIDS-like branching heuristic incorporates random tie-breaking, meaning that when multiple variables have the highest activity scores, one is chosen randomly for branching. 

By contrast to Li et al.'s reset policy, the full reset policy we propose in this paper randomize the activity scores of all variables upon reset, thereby imposing a random branching sequence. Conversely, Li et al.'s solver resets all variable activities to zero, and due to random tie-breaking, the branching order is also randomized. Indeed, the likelihood of both solvers producing the same branching sequence remains equal.

\begin{theorem}\label{thm:li} {\bf Li et al.'s Theorem}~\cite{li2020complexitytheoretic}:
The class of pitfall formulas (from Theorem~\ref{thm:vinyals20hard}) can be solved in polynomial time by CDCL with VSIDS and full resets, except with exponentially small probability.
\end{theorem}

The following corollary follows trivially from Theorem~\ref{thm:vinyals20hard} and  Theorem~\ref{thm:li}.

\begin{corollary}
CDCL solvers with VSIDS-like branching heuristics and traditional restarts is exponentially weaker than CDCL solvers with VSIDS-like branching heuristics and full resets.
\end{corollary}

A natural question is then how do CDCL solvers with VSIDS and partial resets compare with same configuration solvers with full resets. We answer this question below. 
% Are they also stronger than VSIDS-like solvers with restarts?

\begin{observation}\label{obs:partial_upper}
    Pitfall formulas can be solved in polynomial time by CDCL solvers with VSIDS-like branching heuristics and partial reset policies which preserve the order of $O(1)$ of highest active variables upon reset. Further, if $O(log(n))$ variables with highest activities are preserved across reset boundaries, we get a quasi-polynomial upper bound.
\end{observation}

This observation holds true due to the inherent structure of Pitfall formulas, which can be categorized into two components: a hard part (requiring exponential size resolution proofs) and a easy part (having polynomial size resolution proofs). The crux of the lower bound proof (for CDCL solvers with traditional restarts) lies in the likelihood of encountering conflicts primarily within the hard part of the formula. Consequently, the activities of variables within this segment continuously get bumped, making the solver go back to the hard part of the formula even after restarts.

However, solvers with partial reset policies preserving the order of only a constant number of highest activity variables upon reset do not suffer from the above lower bound, these solvers rapidly trigger and resolve conflicts within the hard part of the formula. Since the number of conflicts with a constant number of variables is polynomially bounded, the solver continuously learn all such conflict clauses. Subsequently, even if the top constant number of variables selected for branching are all from the hard part, the solver, armed with learned conflict clauses, will be in a non-conflicting state.

Once this phase concludes, the solver can then strategically utilizes resets to navigate toward a branching sequence to focus on the easy part of the formula. 

\begin{observation}\label{obs:partial_lower}
    Pitfall formulas can only be solved in exponential time by CDCL solvers with VSIDS-like branching heuristics and partial reset policies which preserve the order of $\Omega(n)$ of highest active variables upon reset. 
\end{observation}

The above observation follows directly from the proof of Vinyals' theorem. Consider the extreme case where the order of all variables are preserved after reset, then the solver is really just doing restarts, the solver then suffers from the same lower bound for the restarts solvers considered by Vinyals.

\begin{corollary}
    There exists a $k$, where $0 < k\le n$, and classes of formulas such that CDCL solvers and with VSIDS-like branching heuristics with restart are exponentially weaker than CDCL solvers and with partial resets preserving the order of variables with top $k$ highest activities upon resets.
\end{corollary}

\begin{corollary}
    There exists a $k$, where $0 < k\le n$, and classes of formulas such that CDCL solvers with VSIDS-like branching heuristics and with full reset are exponentially stronger than CDCL solvers with VSIDS-like branching heuristics and with partial resets preserving the order of variables with top $k$ highest activities upon resets.
\end{corollary}

\noindent The above corollaries follow directly from Observation~\ref{obs:partial_upper} and Observation~\ref{obs:partial_lower}.

\begin{figure}[t]
\centering
{\includegraphics[width=0.7\linewidth]{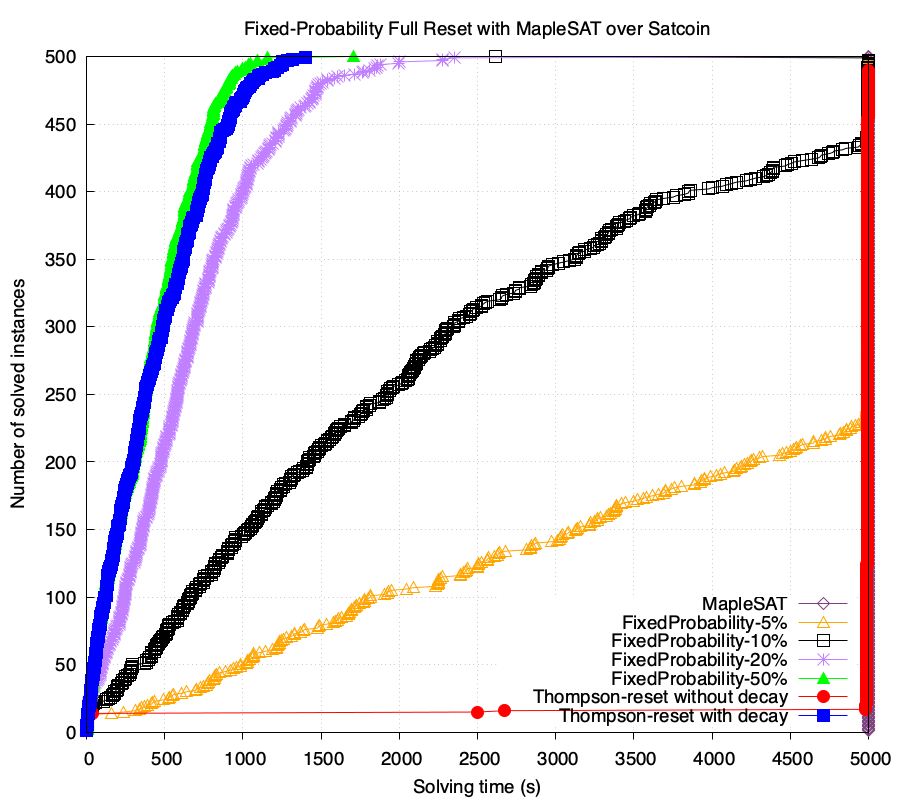}}
\caption{This cactus plot shows the performance of baseline MapleSAT solver vs. fixed probability full reset policies (fix5 - fix50) vs. RL-based full reset on Satcoin instances.}
\label{fig:fix_freq_satcoin} 
\end{figure}

\begin{figure}[t]
\centering
{\includegraphics[width=0.7\linewidth]{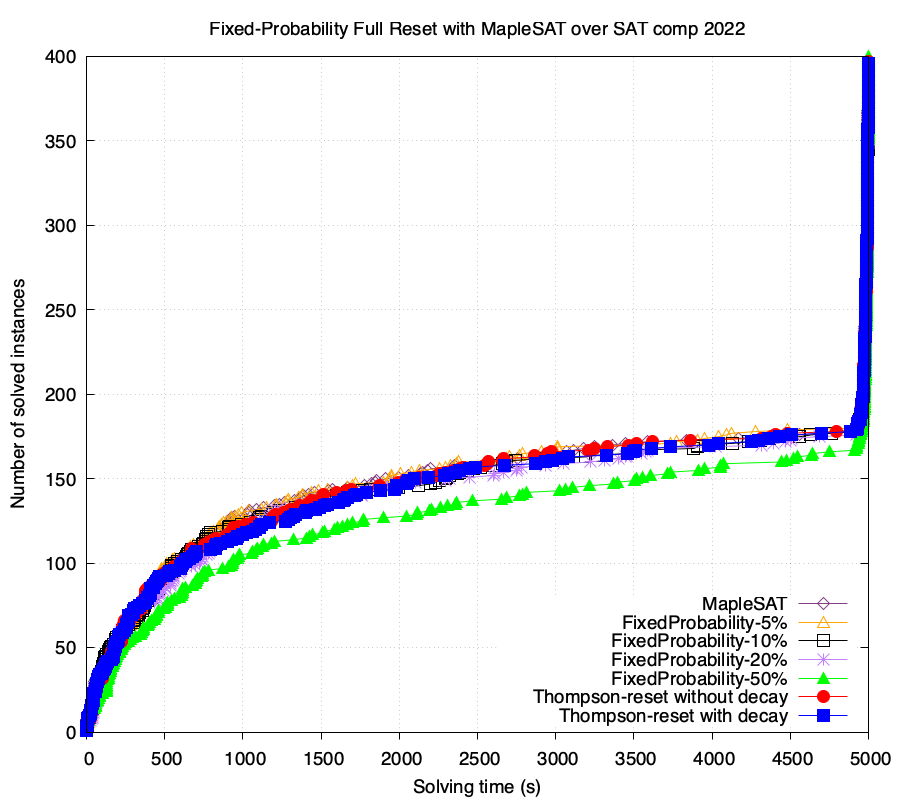}}
\caption{This figure shows that increasing the probability for triggering reset decreases the number of solved Main Track instances. The behavior of baseline Kissat vs. the one with fixed probability full reset policies is similar.
}
\label{fig:fix_freq_main} 
\end{figure}

\subsection{Bounds with Decaying Shape Parameters}

As mentioned above, we apply a decay parameter to the shape parameters of our Thompson sampling based RL-technique for reset invocation. One reason to apply a decay to the shape parameters used in Thompson Sampling is to adapt to the changes in the underlying search space as the solver makes progress in solving an input formula. Here, we present some mathematical consequences of applying decay to the shape parameters of our Thompson sampling based method.

First, note that the values of the shape parameters are upper-bounded by the sum of the geometric series $1+d+d^2 +d^3... = \frac{1}{1-d}$, which is a constant. Therefore, for small values of \(d\), the shape parameters stay at low values at all times.

Also recall the formulas for the mean and variance for the beta distribution: if $X \sim$ beta$(\alpha, \beta)$, then the mean of $X$ is
    $$E[X] = \frac{\alpha}{\alpha + \beta}$$

and the variance of $X$ is 
    $$Var[X] = \frac{\alpha\beta}{(\alpha+\beta)^2(\alpha+\beta+1)}$$

Without a decay, the shape parameters $\alpha$ and $\beta$ may get arbitrarily large and become unbounded. When shape parameters for Thompson Sampling are large and unbounded, the variance when sampling from the underlying beta distribution becomes very small, leading to the following simple observation.

\begin{observation}
\label{obs:beta_var}
$\lim_{\alpha\to\infty} Var[X] \rightarrow 0$ and $\lim_{\beta\to\infty} Var[X] \to 0$.
\end{observation}

This means that the solver would keep picking the same arm, even if the arm fails to perform. A constant change in the values of the shape parameters would not change the underlying distribution too much, the mean would not change much and the variance would remain low, leading to the following observation. Thus, the Thompson Sampling algorithm becomes less exploratory, at least for very long time intervals.

\begin{observation}
\label{obs:beta_mean_large}
When $\alpha >> \beta$ and $\alpha\to\infty$, $\frac{\alpha}{\alpha+\beta}$ - $\frac{\alpha}{\alpha+\beta+1} \to 0$.
\end{observation}

On the other hand, when we apply a decay to the shape parameters, the shape parameters are upper bounded by a constant. Applying a positive constant change in the values of the shape parameters would change the mean by at least a constant, thus making the algorithm more exploratory, leading to the following observation.

\begin{observation}
\label{obs:beta_mean_constant}
For $\alpha, \beta < c$ where $c$ is a positive constant, $\frac{\alpha}{\alpha+\beta}$ - $\frac{\alpha}{\alpha+\beta+1} > \frac{1}{4c+2}$. 
\end{observation}

\begin{table}[t]
\begin{tabular}{l|lll|}
% \cline{2-4}
                                                                             \cline{2-4} 
                                            & \multicolumn{1}{l|}{SAT comp 2022} & \multicolumn{1}{l|}{SAT comp 2023} & Satcoin \\ \hline
\multicolumn{1}{|l|}{Kissat3.0.0}           & \multicolumn{1}{l|}{272}           & \multicolumn{1}{l|}{178}           & 0       \\ \hline
\multicolumn{1}{|l|}{Kissat3.0.0\_Thompson} & \multicolumn{1}{l|}{\bf276}           & \multicolumn{1}{l|}{234}           & 360     \\ \hline
\multicolumn{1}{|l|}{Kissat3.1.0}           & \multicolumn{1}{l|}{270}           & \multicolumn{1}{l|}{250}           & \bf500     \\ \hline
\multicolumn{1}{|l|}{Kissat3.1.0\_Thompson} & \multicolumn{1}{l|}{270}           & \multicolumn{1}{l|}{\bf253}           & \bf{500}     \\ \hline
\end{tabular}
\caption{Number of solved instances by each Kissat variant. The bold numbers highlight the best performance achieved by each solver in the corresponding benchmark.}
%\caption{Number of solved instances by each Kissat variant. The bold numbers highlight the best performance achieved by each solver in the corresponding benchmark. The results show that Kissat3.0.0\_Thompson performed the best in SAT comp 2022 with 276 instances solved, while Kissat3.1.0\_Thompson had the highest number of solved instances (253) in SAT comp 2023.}
\label{tbl:kissat}
\end{table}

\begin{table}[t]
\begin{tabular}{l|lll|}
 \cline{2-4} 
                                         & \multicolumn{1}{l|}{SAT comp 2022} & \multicolumn{1}{l|}{SAT comp 2023} & Satcoin      \\ \hline
\multicolumn{1}{|l|}{MapleSAT}           & \multicolumn{1}{l|}{\textbf{178}}  & \multicolumn{1}{l|}{144}           & 13           \\ \hline
\multicolumn{1}{|l|}{MapleSAT\_Thompson} & \multicolumn{1}{l|}{\textbf{178}}  & \multicolumn{1}{l|}{\textbf{164}}  & \textbf{500} \\ \hline
\end{tabular}
\caption{Number of solved instances by MapleSAT and MapleSAT\_Thompson. The bold numbers highlight the best performance achieved by each solver in the corresponding benchmark.}
% \caption{Number of solved instances by MapleSAT and MapleSAT\_Thompson. The bold numbers highlight the best performance achieved by each solver in the corresponding benchmark. The results demonstrate that MapleSAT and MapleSAT\_Thompson achieved comparable performance in SAT comp 2022. However, MapleSAT\_Thompson outperformed MapleSAT in SAT comp 2023 and Satcoin, solving 164 and 500 instances respectively, compared to MapleSAT's 144 instances in SAT comp 2023 and 13 instances in Satcoin.}
\label{tbl:MapleSAT}
\end{table}

\begin{table}[t]
\begin{tabular}{l|lll|}
                                 \cline{2-4} 
                                              & \multicolumn{1}{l|}{SAT comp 2022} & \multicolumn{1}{l|}{SAT comp 2023} & Satcoin      \\ \hline
\multicolumn{1}{|l|}{CaDiCaL}                 & \multicolumn{1}{l|}{\textbf{277}}  & \multicolumn{1}{l|}{253}           & 489          \\ \hline
\multicolumn{1}{|l|}{CaDiCal\_Thompson}       & \multicolumn{1}{l|}{263}           & \multicolumn{1}{l|}{240}           & \textbf{500} \\ \hline
\multicolumn{1}{|l|}{SBVA\_Cadical}           & \multicolumn{1}{l|}{248}     & \multicolumn{1}{l|}{\textbf{272}}     & 480          \\ \hline
\multicolumn{1}{|l|}{SBVA\_Cadical\_Thompson} & \multicolumn{1}{l|}{238}              & \multicolumn{1}{l|}{257}              &           \bf{500}   \\ \hline
\end{tabular}
\caption{Number of solved instances by each CaDiCal variant. The bold numbers highlight the best performance achieved by each solver in the corresponding benchmark.}
%\caption{Number of solved instances by each CaDiCal variant. The bold numbers highlight the best performance achieved by each solver in the corresponding benchmark. RL reset policy improve the performance of CaDiCal and SBVA\_Cadical on Satcoin, however underperform on SAT comp instances.}
\label{tbl:cadical}
\end{table}

\section{Experimental Results}

In this Section, we report on the extensive experimental evaluation we performed to test the above-described reset policies as implemented in CaDiCal, SBVA Cadical, Kissat and MapleSAT solvers. For Kissat we use two versions of Kissat, the latest version Kissat3.1.0 as well as the previous version Kissat3.0.0. We refer the reader to Table~\ref{tbl:kissat},~\ref{tbl:MapleSAT} and~\ref{tbl:cadical} containing a summary of total number of solved instances for our selected benchmarks. 

\subsection{Experimental Setup}
All experiments were conducted on an Intel E5-2683 v4 Broadwell @ 2.1GHz CPUs running Linux
3.10.0-1160.88.1.el7.x86 64 (Digital Research Alliance of
Canada), with a timeout of 5000 seconds of CPU time. The timeout is standard in the SAT community~\cite{SATcomp}.

Since there is randomness involved in our Thompson-based reset solvers (for sampling from beta distribution), we use the same default random seeds as the baseline solvers. We use 0.8 as the decay factor for both computing the EMA\_\rwglr{} as well as decaying the shape parameters used in Thompson sampling. For SW-UCB, we use the window size $\tau$ of 30 and the exploration constant $c$ of 0.2. 

\subsection{Benchmarks}

We use the following three sets of benchmarks for evaluating our solvers.

\begin{enumerate}
    \item \textbf{Satcoin instances~\cite{manthey2018satcoin}:} These instances are derived from Bitcoin mining problems. Satcoin instances are optimization problems that involve finding a nonce value that, when combined with a block header, results in a hash value that is lower than a given difficulty target. By varying the difficulty target, one can generate Satcoin instances of various hardness and perform scaling studies.
    
    \item \textbf{Main Track instances from SAT Competition 2022 and 2023~\cite{SATcomp}:} The Main Track instances are a set of problems used in SAT Competition 2022 and 2023. The Main Track instances are designed to be representative of real-world SAT problems and cover a wide range of applications, including verification, planning, and scheduling. The instances are a mix of easy, medium, and hard instances. The Main Track instances are a widely recognized benchmark for evaluating the performance of SAT solvers.
\end{enumerate}

\subsection{Results of Full Reset with Fixed Probabilities}
\label{sec:fix_prob}
By performing experiments on Satcoin instances using fixed probability reset policy, the following observations were made for MapleSAT and Kissat3.0.0 (see Figure~\ref{fig:fix_freq_satcoin}): with the fixed probability reset policy, we solve significantly more Satcoin instances than the baseline solvers which can barely solve any of these instances. Even with 5\% probability of reset, the reset based CDCL solver solves significantly more Satcoin instances within 5000 seconds timeout limit. We observed that as we increase the probability of triggering a reset, MapleSAT solved all 500 out of 500 Satcoin instances at 20\% fixed probability reset. When the probability goes up from 20\% to 50\%, the time used to solve all 500 instances continues to reduce.

However, by using fixed probability reset policy on Main Track instances from SAT competition 2022 and 2023, we saw very different behavior of the reset based CDCL solvers (see Figure~\ref{fig:fix_freq_main}): when we have a relatively small reset probability, the reset policy's performance can match the baseline solver. For example for SAT competition 2022, 5\% reset probability MapleSAT solver can solve 178 instances, which is the same number of solved instances by the base solver. However, as we increase the reset probability, the performance on Main Track instances degrades significantly. With 50\% reset probability MapleSAT only solves 166 instances.

The experiments with fixed probability reset policy for MapleSAT and Kissat3.0.0 demonstrate the power of reset in solving Satcoin instances, while on the other hand an overly resetting solver leads to poor performance on Main Track instances. This dilemma motivates the RL-based reset policies aiming at dynamically and adaptively choosing when to invoke reset during a CDCL solver's run.

\subsection{Results of Thompson Full Reset}
During our experiments with Thompson sampling based reset polices, we observed that the Thompson sampling algorithm was strongly biased by the initial warm-up of the solvers. To be more specific the solver is biased to not do any reset and makes its decision to not do a reset based on the outcomes from the initial warm-up stage of the search. Due to a need of focus on more recent outcomes, we apply a decay to alpha and beta used in Thompson sampling.

By experimenting Thompson reset in MapleSAT and Kissat3.0.0 on Satcoin and SAT Competition 2022 and 2023 instances, the following results are obtained (see Figure~\ref{fig:fix_freq_satcoin} and Figure~\ref{fig:fix_freq_main}): For Satcoin instances, Thompson reset without decaying alpha and beta values barely solves any instances. This is due to the warm-up bias mentioned above. In the early stage of the search, the RL model learns not to trigger reset frequently. This bias was taken to the later stage of the search, which significantly reduced the number of resets being carried out. As we see in Section~\ref{sec:fix_prob}, Satcoin instances can be solved efficiently with a higher probability of reset. Therefore, it becomes reasonable for this version of RL reset to perform poorly. With decaying alpha and beta values, the RL model focuses more on the outcomes of recent observations. Thus, more resets are triggered when a recent search has poor \rwglr{}. With decaying alpha and beta values, we not only solve all 500 instances but also perform almost as well as the 50\% fixed probability reset solver. For SAT comp 2022 and 2023 instances, solvers with Thompson reset and decaying parameters strictly improve the performances of their respective baseline solvers.

\subsection{Results of Partial Reset}
We observe that when implementing a partial reset strategy, the optimum value of $k$, which determines how many of the top $k$ variables should be preserved post reset, is not clear. For main track instances, among $k = 5, 10, 20 \text{ and } 30$, $k = 5$ has the best performance, but yet $k = 10$ performs worse than $k = 30$. On a high level, the performance for partial reset is only minorly different from the full reset version.

\subsection{Results of UCB Reset}
During our extensive evaluations for SAT comp 2022, 2023, and Satcoin, we applied both Thompson sampling-based reset policies and UCB-based reset policies. We observe that while both approaches exhibited are effective, the Thompson sampling method consistently outperformed the UCB approach. This underscores the robustness of Thompson sampling.

\subsection{Overall Evaluation of Thompson reset}
We implemented Thompson Sampling-based reset policies over five baseline solvers, CaDiCal, SBVA Cadical, Kissat3.0.0, Kissat3.1.0 and MapleSAT and evaluated their performance over SAT comp 2022, 2023 and Satcoin benchmarks (see Table~\ref{tbl:kissat}, Table~\ref{tbl:MapleSAT} and Table~\ref{tbl:cadical}). 
The evaluation highlights the specific benchmarks where the Thompson variant outperformed its baseline counterpart. For Kissat solvers, Kissat3.0.0\_Thompson notably excelled in SAT comp 2022 by solving 276 instances, surpassing the baseline Kissat3.0.0. And in SAT comp 2023, Kissat3.1.0\_Thompson also outperforms its baseline. For MapleSAT, MapleSAT\_Thompson demonstrated superior performance in SAT comp 2023 and Satcoin, solving 164 instances in the former and an impressive 500 instances in the latter, outperforming MapleSAT in both tracks. For the SAT comp 2022, the Thompson variant of MapleSAT is as competitive as its baseline. For CaDiCal however, our RL policy slows down baseline solvers' performance over SAT comp instances, but CaDiCal\_Thompson and SBVA\_Cadical\_Thompson showcased enhanced performance in Satcoin.

\section{Related Work}

There has been considerable theoretical and empirical work on restarts. Liang et al.~\cite{liang2018machine} provide a nice overview of the empirical work on restarts, while the paper by Li et al.~\cite{li2020complexitytheoretic} gives a thorough overview of the proof complexity of restarts. Below we focus mostly on reset policies, and contrast with restart methods as appropriate. 

\subsection{Empirical Work on Reset Policies}
Chaff~\cite{moskewicz2001chaff} was the first CDCL solver to implement a restart policy. To the best of our knowledge, CaDiCal~\cite{BiereFazekasFleuryHeisinger-SAT-Competition-2020-solvers} is the first CDCL solver with the idea of rephasing, where the phase value of variables is reset on certain intervals during the run of the solver. While many solver developers have experimented with reset policies, we are not aware of a thorough and systematic study of such techniques. To the extent that we know, we are the first to explore RL-based full and partial reset policies.

\subsection{Theoretical Work on Reset Policies}
The first paper to discuss restarts in the context of DPLL SAT solvers was by Gomes and Selman~\cite{gomes2000heavy}. They proposed an explanation for the power of restarts popularly referred to as ``heavy-tailed explanation of restarts''. Their explanation relies on the observation that the runtime of randomized DPLL SAT solvers on satisfiable instances, when invoked with different random seeds, exhibits a heavy-tailed distribution. This means that the probability of the solver exhibiting a long runtime on a given input and random seed is non-negligible. However, because of the heavy-tailed distribution of solver runtimes, it is likely that the solver may run quickly on the given input for a different random seed. This observation was the motivation for the original proposal of the restart heuristic in DPLL SAT solvers by Gomes and Selman~\cite{gomes2000heavy}.

\subsection{Use of Reinforcement Learning in Solvers}
% John: may add neuroback
There is considerable work on the use of RL techniques for switching between solver heuristics or in the context of proof rule sequencing and selection. The earliest work for using RL to switch between heuristics that we are aware of is by Lougdakis et al.~\cite{lagoudakis2001learning}. In their work, they switch between different branching heuristics. Liang et al.~\cite{liang2016learning} were among the first to model branching as an MAB problem, where each arm represents a variable to be branched on. Liang et al.~\cite{liang2018machine} also developed an RL technique for switching between restart policies. Another example of the use of RL techniques to switch between branching heuristics is the Kissat-MAB-Hywalk solver~\cite{zheng2022combining}. In recent years, Kurin et al. \cite{GQSAT2019} and Cameron et al. \cite{mcfs_2022} have attempted to learn a deep RL branching policy that maps from the solver state to a branching variable. Despite the novelty, their performance has not yet reached the efficiency level demonstrated by modern SAT solvers. 

Our work differs from these previous methods in the following important ways. First, we adapt Thompson sampling to our setting to solve the problem of switching between reset heuristics, a technique that does not seem to have been explored in the solver context previously. Second, we use a shaping mechanism to help our learning agent to explore more than it otherwise would. Third, we develop our technique for the setting of reset policies. Another point of differentiation is that we are not aware of previous work that thoroughly and systematically explores a variety of reset policies in multiple SOTA solvers over a comprehensive set of benchmark suites.

\section{Conclusions and Future Work}

In this paper, we revisit the idea of reset, a variant of restart, at certain intervals during the run of the solver. Specifically, we study a range of reset policies and devise an RL based technique to switch between restart and reset in the baseline solvers. We observe that fixed probability full reset (i.e., randomizing the activity of all variables) performs dramatically well for Satcoin instances for MapleSAT and Kissat3.0.0. However, the same technique is somewhat weaker on the SAT competition 2022 and 2023 main track instances compared to the corresponding baselines. This behavior of reset motivates the design of reinforcement learning policies that adaptively and in an online fashion chooses to invoke a reset based on the success of previous invocations, as measured by an EMA over a metric we call restart window global learning rate. 

We implement our techniques in five state-of-the-art baseline solvers, namely, CaDiCal, SBVA\_Cadical, Kissat3.0.0, Kissat3.1.0 and MapleSAT. Via extensive experimentation, we show that our RL-based full reset techniques vastly outperform the baseline (MapleSAT and Kissat3.0.0) on Satcoin benchmarks, and improve upon the baselines (MapleSAT, Kissat3.0.0 and Kissat3.1.0) over the SAT competition 2022 and 2023 Main Track instances. We also show minor improvements for CaDiCal and SBVA\_Cadical over Satcoin benchmarks. We also introduce the concept of a partial reset, where at least a constant number of variable activities are retained across reset boundaries. Building on previous results, we show that there is an exponential separation between $O(1)$ vs. $\Omega(n)$-length partial resets.

% We model the problem of whether to restart with resetting activities or restart without resetting activities as a multi-armed bandit (MAB) problem with two arms, where pulling arm one corresponds to restarting with resetting activities and arm two being restarting without resetting activities. We then adapt the Thompson sampling algorithm for MAB problems in our modeling, where we define the shape parameters (successes and failures) through a mechanism using the idea of learning rates. We further propose a technique which applies a decay to the shape parameters to capture the performance of each arm under constant changes in search space as solvers learn and delete clauses. 

% We experimentally evaluated our techniques over two state-of-the-art baseline solvers, namely maplesat and kissat, where we show that our techniques dramatically improved the performance of the baseline solvers over a class of crypto benchmarks derived from bitcoin mining problems, and yet the solvers performance remains competitive over SAT competition instances. Lastly, we explore the idea of partially resetting activities, and we show that partial resetting improved the performance of our solvers over SAT competition instances, while maintaining exceptional performance over the bitcoin mining instances. 

\bibliography{CDCL_RL_reset}

\end{document}